\documentclass[conference]{IEEEtran}
\usepackage{amsthm, amsmath}
\usepackage{float}
\usepackage{xcolor}
\usepackage{array}
\usepackage{booktabs}
\usepackage{graphicx}
\usepackage{subcaption}
\usepackage{threeparttable}
\usepackage[justification=centering]{caption}
\usepackage[font=small,labelfont=bf]{caption}
\usepackage[hyphens]{url}
\usepackage[bookmarks=false]{hyperref}
\hypersetup{breaklinks=true}
\usepackage{cite}
\usepackage[export]{adjustbox}
\usepackage{tabularx}
\usepackage[utf8]{inputenc}
\usepackage{comment}

\usepackage{color}
\usepackage{xcolor,colortbl}
\usepackage{multicol}
\usepackage{multirow}
\usepackage{hhline}
\usepackage{makecell}

\definecolor{codegreen}{rgb}{0,0.6,0}
\definecolor{codegray}{rgb}{0.5,0.5,0.5}
\definecolor{codepurple}{rgb}{0.58,0,0.82}
\definecolor{backcolour}{rgb}{0.95,0.95,0.92}
\definecolor{LightBlue}{rgb}{0.83, 0.91, 1}
\definecolor{LightGreen}{rgb}{0.8, 1, 0.8}
\definecolor{LightPink}{rgb}{1, 0.8, 0.88}
\definecolor{LightYellow}{rgb}{1, 1, 0.6}
\definecolor{light-gray}{gray}{0.9}
 
\usepackage{listings}
\lstdefinestyle{mystyle}{
    backgroundcolor=\color{backcolour},   
    commentstyle=\color{codegreen},
    keywordstyle=\color{magenta},
    numberstyle=\tiny\color{codegray},
    stringstyle=\color{codepurple},
    basicstyle=\ttfamily\footnotesize,
    breakatwhitespace=false,         
    breaklines=true,                 
    captionpos=b,                    
    keepspaces=true,                 
    numbers=left,                    
    numbersep=5pt,                  
    showspaces=false,                
    showstringspaces=false,
    showtabs=false,                  
    tabsize=2
}

\begin{document}

\title{Compute RAMs: Adaptable Compute and Storage Blocks for DL-Optimized FPGAs
\vspace{-0.3cm}}


\author{
\IEEEauthorblockN{Aman Arora\IEEEauthorrefmark{1}, 
Bagus Hanindhito\IEEEauthorrefmark{2},
Lizy K. John\IEEEauthorrefmark{3}
}
\IEEEauthorblockA{
\textit{The University of Texas at Austin}\\
\IEEEauthorrefmark{1}aman.kbm@utexas.edu,
\IEEEauthorrefmark{2}hanindhito@bagus.my.id,
\IEEEauthorrefmark{3}ljohn@ece.utexas.edu
}
\vspace{-5ex}
}




\begin{figure*}[!t]
\begin{large}
\textcopyright 2021 IEEE. Personal use of this material is permitted.  Permission from IEEE must be obtained for all other uses, in any current or future media, including reprinting/republishing this material for advertising or promotional purposes, creating new collective works, for resale or redistribution to servers or lists, or reuse of any copyrighted component of this work in other works.
\newline
\newline
\newline
\newline
This work was submitted to \textbf{IEEE Signal Processing Society's ASILOMAR Conference on Signals, Systems and Computers} on the midnight of May 8, 2021. This paper has been accepted by the conference (notification received on July 9, 2021) and will be published in the proceedings of the conference. The conference dates are Oct 31 - Nov 3, 2021. Copyright may be transferred without notice, after which this version may no longer be accessible. 
\end{large}
\end{figure*}
\pagebreak

\maketitle

\begin{abstract}

The configurable building blocks of current FPGAs --- Logic blocks (LBs), Digital Signal Processing (DSP) slices, and Block RAMs (BRAMs) --- make them efficient hardware accelerators for the rapid-changing world of Deep Learning (DL).
Communication between these blocks happens through an interconnect fabric consisting of switching elements spread throughout the FPGA.
In this paper, a new block, Compute RAM, is proposed. Compute RAMs provide highly-parallel processing-in-memory (PIM) by combining computation and storage capabilities in one block. 
Compute RAMs can be integrated in the FPGA fabric just like the existing FPGA blocks and provide two modes of operation (storage or compute) that can be dynamically chosen.
They reduce power consumption by reducing data movement, provide adaptable precision support, and increase the compute density of FPGAs.
In our evaluation of addition, multiplication and dot-product operations across multiple data precisions (int4, int8 and bfloat16), we observe an average savings of 80\% in energy consumption, and an improvement in execution time ranging from 20\% to 80\%. 
Adding Compute RAMs can benefit non-DL applications as well, and make FPGAs more efficient, flexible, and performant accelerators.
\end{abstract}

\IEEEpeerreviewmaketitle

\section{Introduction} \label{introduction}


Deep Learning (DL) has become ubiquitous in today's world. The ever-increasing computational demands of DL applications has triggered an explosion of hardware acceleration alternatives, ranging from ASICs to GPUs to FPGAs. FPGAs are well-suited to the evolving needs of DL applications because they provide customizable hardware with massive parallelism and energy efficiency. 

FPGAs contain fine-grained programmable logic (e.g., LBs), fixed-function math units (e.g., DSP slices), and embedded memory structures (e.g., BRAMs) that can be connected by a configurable interconnection fabric. These building blocks are very generic, making FPGAs a great solution to design various accelerators, but this flexibility, unfortunately, limits the performance we can achieve with FPGAs for DL applications. In recent years, DL-optimized FPGA architectures have been proposed and deployed, such as adding vector processors \cite{xilinx_versal_ai} and integrating DL-specific blocks \cite{langhammer2021stratix}\cite{tensor_slice_paper} on the FPGA chip. Most FPGA vendors have added support for smaller,  DL-friendly  precisions (e.g., 8-bit fixed-point (\texttt{int8}) and \texttt{bfloat16} \cite{bfloat16}) in  DSP slices.


Even so, there are still some limitations in current FPGA architectures. Separation of compute units (LBs and DSPs) from storage units (BRAMs) leads to a lot of data movement to feed the compute units with input data and to store the outputs back to the storage units. This is exacerbated for DL applications because of the math-intensive nature of operations involved in them. This data movement, although on-chip, is expensive in terms of power consumption because the movement happens through the FPGA interconnect which comprises of numerous switches instead of hard connected wires. This flexible but inefficient interconnect also leads to slower frequencies for designs on FPGAs (typically 3-4x lower than ASICs \cite{fpga_asic_gap}).

BRAMs on FPGAs support a limited set of heights and widths. For example, BRAMs in Intel Agilex \cite{intel_agilex} are 20 Kilobits in size and can be configured as 512x40, 1024x20 and 2048x10 bits, with only 1 or 2 read and write ports. Owing to the parallel nature of DL applications, it is common to process thousands of bits of data together. It is preferred for users to split the data over multiple BRAMs for higher bandwidth, which leads to only a few rows of each block are utilized. 

Another limitation is the limited number of precisions supported by the DSP slices. For example, DSP slices in Intel Agilex FPGAs \cite{intel_agilex} support multiplication and MAC (multiply-accumulate) operations in 9x9, 27x27, 18x19 fixed-point and 16-bit or 32-bit floating-point precisions. Although DSP slices have become more complex over the years to support more precisions, the precision requirements change rapidly, especially in the world of DL. Users have to implement math units on LBs instead, reducing the number of LBs available for other purposes and leaving DSPs unused.



\begin{figure}[t!]
\centering
\includegraphics[width=0.9\linewidth]{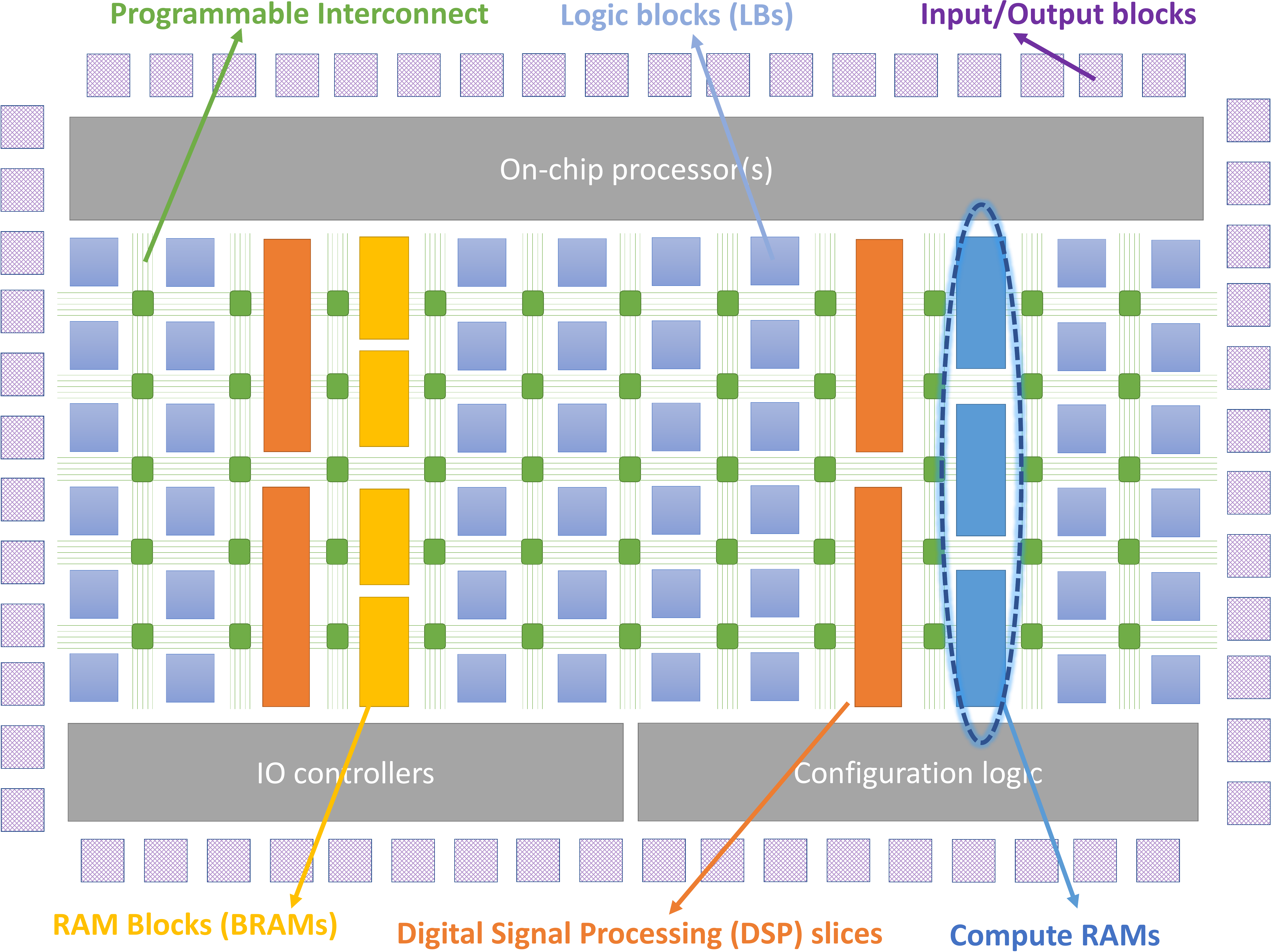}
\caption{An FPGA with the proposed Compute RAM blocks}
\label{fig:fpga_with_cram}
\vspace{-2mm}
\end{figure}

In this paper, we propose adding a new type of block, called Compute RAM, to FPGAs. A Compute RAM block enables computation within the RAM array, without transferring the data in or out of it. The implementation of these blocks is based on an emerging Logic-in-Memory SRAM prototype by Jeloka et. al. \cite{supreet_logic_in_memory}. Using this technology and its extensions (bit-line computing \cite{compute_cache} and bit-serial arithmetic \cite{neural_cache}), processing-in-memory engines can be designed. For Compute RAMs, we add components to such SRAMs to integrate them into the FPGA fabric and make them configurable. Compute RAMs can be programmed during FPGA configuration time or during run-time. 
The user can perform math in any precision using Compute RAMs. 
Every column of the memory performs the same operation simultaneously, resulting in massive parallelism and high throughput.
Since there is no need to move the data in and out of the block, the energy efficiency improves drastically. A reduced dependency on FPGA interconnect also means using Compute RAMs leads to faster frequencies. On the top of these advantages, the blocks can still be used as pure storage blocks as needed. A block diagram of an FPGA with Compute RAMs is shown in Figure \ref{fig:fpga_with_cram}.


Here is the summary of the contributions of this paper:
\begin{enumerate}
    \item Propose adding blocks called Compute RAMs to FPGAs 
    \item Describe the architecture of Compute RAM blocks and their various features
    \item Demonstrate the benefit of deploying Compute RAMs for common DL operations
\end{enumerate}


\section{Related Work and Background} \label{section:related_work}

\subsection{DL-Optimized FPGAs}

In recent years, many DL-specific modifications to FPGA architecture have been deployed by the industry. Xilinx Versal family \cite{xilinx_versal_ai} adds specialized vector processors for DL acceleration. Intel's Stratix 10 NX FPGAs integrate in-fabric AI tensor blocks \cite{langhammer2021stratix}. Achronix Speedster7t FPGAs \cite{achronix_mlp} have embedded machine learning processor (MLP) blocks that have an array of multipliers, an adder tree and accumulators. The FlexLogix nnMAX \cite{flexlogix_nnmax} inference IP also contains hard blocks to perform convolutions. Native support for \texttt{fp16} and \texttt{bfloat16} data types in DSP slices has also been added to recent FPGAs.

There have also been a number of academic research proposals to optimize FPGA architectures for DL. Eldafrawy et al. \cite{logic_block_mohd} proposed several enhancements to the LB architecture, including adding a shadow multiplier in them. In \cite{embrace_div_dsp}, the Boutros et al. propose enhancing DSP blocks by efficiently supporting low precision multiplications. Rasoulinezhad et al. \cite{pir_dsp} have proposed DSP slice modifications such as including a register file for data reuse and improvements to DSP-DSP interconnect. Arora et al. \cite{tensor_slice_paper} also proposed adding Tensor slices in FPGAs.

To our knowledge, no existing work proposes adding processing-in-memory (PIM) blocks to the FPGA fabric.

\subsection{Processing-In-Memory}

Proposals for Processing-In-Memory (PIM) \cite{pim_workload_perspective} architectures have been around for decades, but the products which implement it can only be seen in the recent years, specifically for DL acceleration. Memristor-based PIM accelerators like ISAAC \cite{isaac} and PRIME \cite{prime} were the early entrants in this field, but recently many digital solutions have been proposed as well, such as FloatPIM \cite{floatpim} which has support for floating point operations. The main limitation of these architectures is integrating them on the same Silicon die that uses a standard CMOS-based process. Only some vertical stacked architectures have been shown to work so far \cite{nature_memristor_cmos}. 

Samsung recently announced a DRAM product called HBM-PIM \cite{samsung_hbm_pim} which integrates compute units onto a High Bandwidth Memory chip. This paradigm is near-memory compute instead of in-memory compute. Mythic AI's Intelligent Processing Unit (IPU) contains tiles that have analog matrix multiplier which uses Flash memory transistors. Their design requires DACs and ADCs for operation.


Jeloka et. al. \cite{supreet_logic_in_memory} showcased a Logic-in-Memory SRAM prototype. Multiple word lines are activated simultaneously and the shared bit-lines can be sensed, effectively performing logical AND and NOR operations on the data stored in the activated rows. By lowering the word-line voltage, data corruption due to multi-row access is prevented. 
Aga et al. \cite{compute_cache} proposed deploying these bit-line computing enabled SRAMs as caches in CPUs to create massively parallel compute engines. They extend the technology to add the capability of operations like compare, NOT, XOR, copy, search, etc. 

\begin{figure*}[hbt!]
\centering
\includegraphics[width=0.9\linewidth]{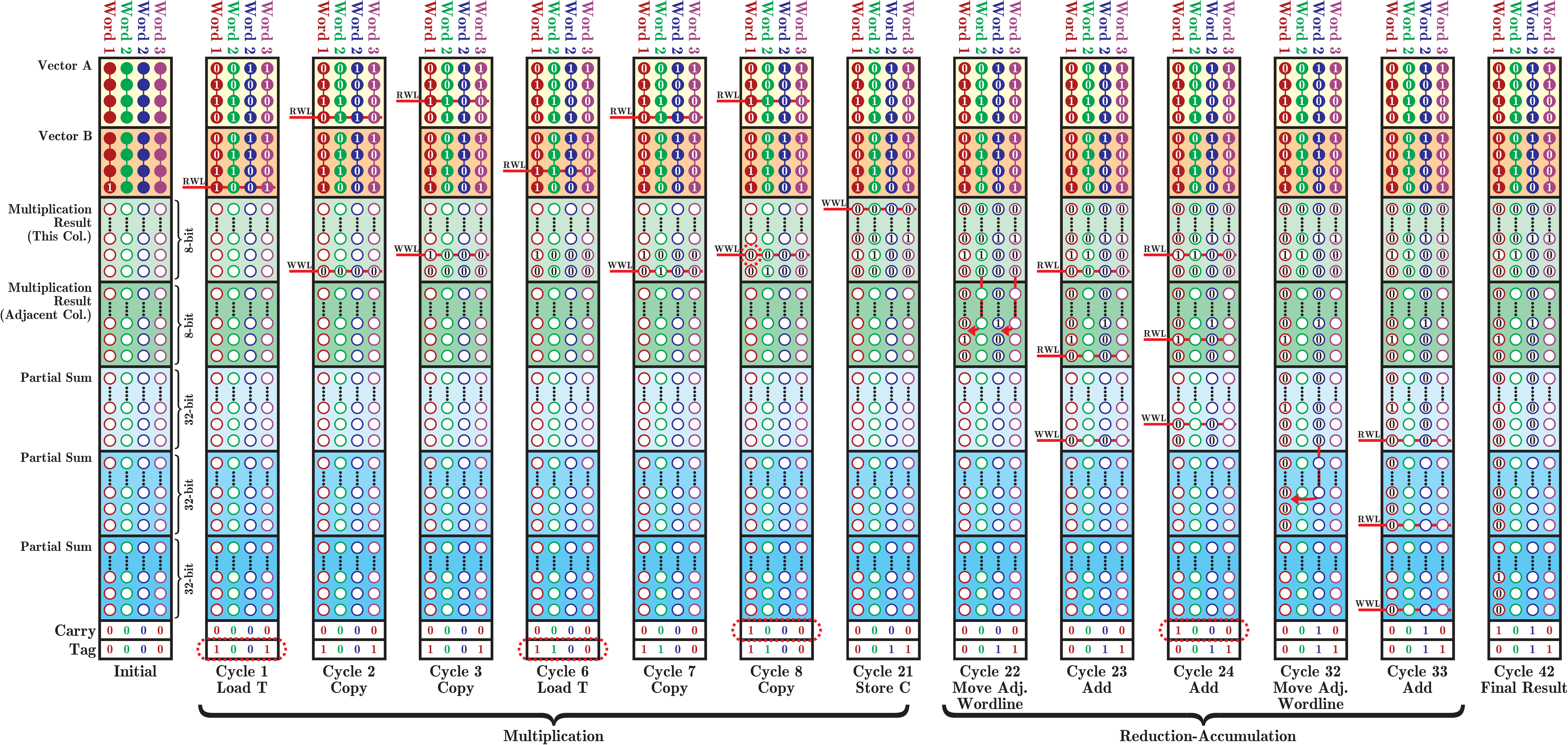}
\caption{Performing a dot product based on the architecture in \cite{neural_cache} (WWL = Write Word Line, RWL = Read Word Line)}
\label{fig:neural_cache}
\vspace{-2mm}
\end{figure*}

Eckert et al. \cite{neural_cache} combine this capability with bit-serial arithmetic, which involves processing one bit of multiple data elements every cycle, instead of processing multiple bits of 1 data element every cycle. Data is stored in a transposed format in the array. That is, the bits of operands are stored in one column (i.e. in multiple word lines). Some logic gates are added near the sense amplifiers of each column (bit-line) to make performing arithmetic operations easier. In the first half of a clock cycle, two bits of operands are read, and logical or arithmetic operations are performed on them. The result is written back into the rows of the same column in the second half of the same clock cycle.
Figure \ref{fig:neural_cache} shows how a dot product operation can be performed using this method.

In this paper, we propose Compute RAMs which extend and enhance this technology for FPGAs. Note that Compute RAMs are not limited to using this technology; any memory technology as that provides bit-level computing can be used to design Compute RAMs.

\vspace{-2mm}

\section{Proposed Architecture: Compute RAM} \label{proposal}

\begin{figure}[hbt!]
\centering
\includegraphics[width=0.9\linewidth]{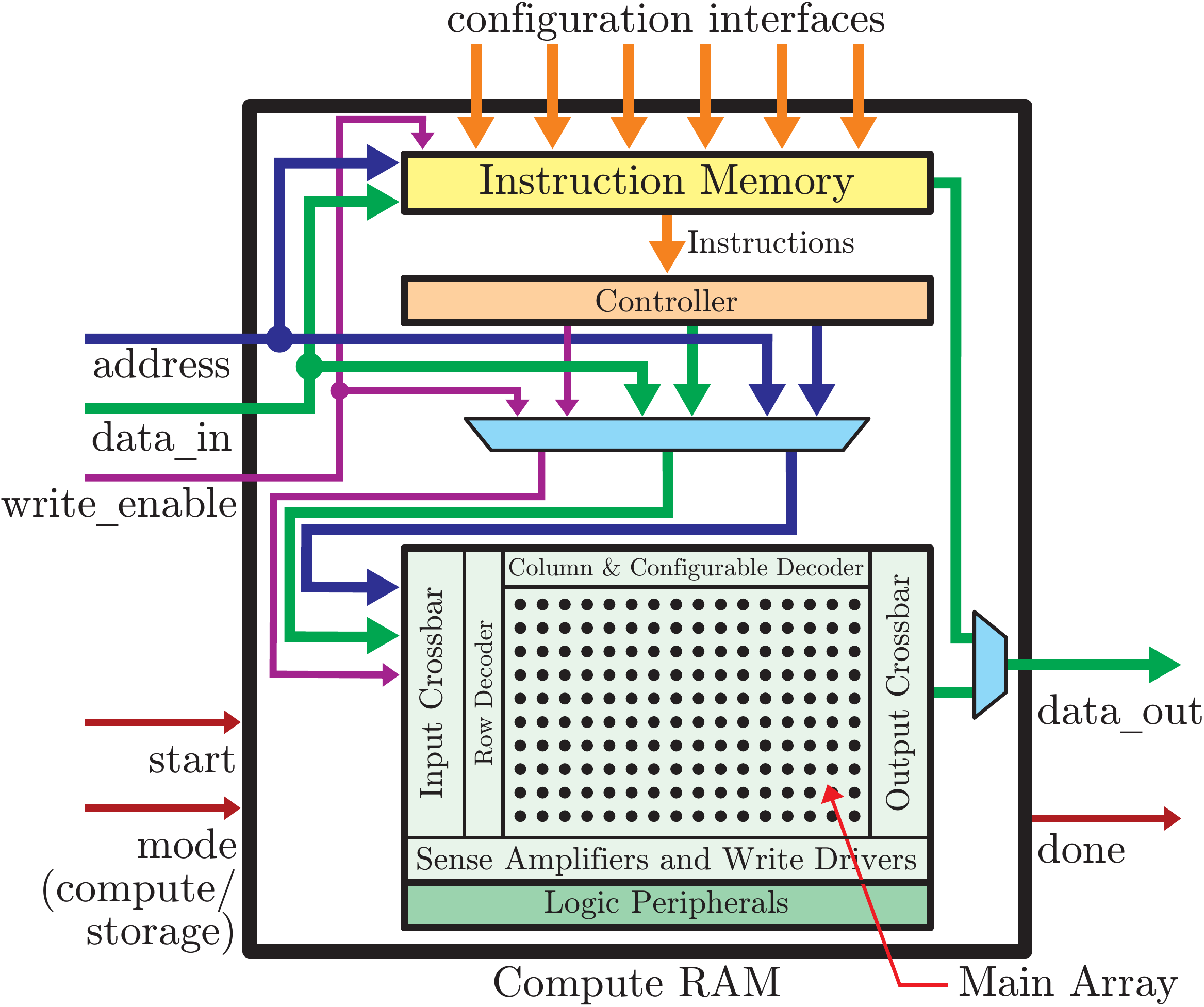}
\caption{Architecture of a Compute RAM block}
\label{fig:compute_ram_arch}
\vspace{-2mm}
\end{figure}

Figure \ref{fig:compute_ram_arch} shows the architecture of the proposed Compute RAM block. The heart of the Compute RAM block is an SRAM array (called the main array) that supports bit-line computing, as prototyped in \cite{supreet_logic_in_memory}. The instruction memory is a small SRAM that contains the instructions for operation to be performed on the data. For example, it could contain the instruction sequence for performing \texttt{int4} (4-bit fixed-point) additions on the data in the array. This software-like mechanism enables users to perform computations with any precision using Compute RAMs.
A controller reads and decodes the instruction sequence stored in the instruction memory. Based on these instructions, it sends commands to the array to perform the computations. This reduces the length of instruction sequences required to perform more complex operations. Logic peripherals, enhanced compared to \cite{neural_cache}, are present near the sense amplifiers and write drivers to perform compute operations. 

\vspace{-2mm}
\vspace{-2mm}

\subsection{Details of Components}

\subsubsection{Main Array}
The main array is an SRAM array that supports logic-in-memory or bit-line computing. It's operation is briefly described in Section \ref{section:related_work}. 
This SRAM array is a drop-in replacement of a typical BRAM on an FPGA. Some additional muxes are added around the array to enable access to its inputs and outputs during the compute mode. FPGAs have additional circuitry to allow for configurable geometries (height and width) \cite{mtj_ram}. The same circuitry can be used here without changing the operation of the Compute RAM. Because the compute operations are done in parallel across columns, to obtain the most parallelism, it is best to configure the Compute RAM as wide as possible and as shallow as possible. We use the same size and geometries as BRAMs in Intel FPGAs. The main array in Compute RAM is 20 Kilobits in size and can be configured in 512x40, 1024x20 and 2048x10 geometries. Both row decoders available in the array (BRAMs are dual ported) are used in the compute mode.

\subsubsection{Instruction Memory}
An SRAM is provided to store the sequence of instructions for the operations to be done on the data in the main array and can be loaded in two ways:
\begin{itemize}
    \item \textbf{At FPGA configuration time.} To allow this, we provide connection for this memory to talk to the FPGA configuration interface. This method can be used when the operation executed by the block is not expected to be changed during execution time. 
    \item \textbf{At execution time.} Sometimes, the operation that will be performed on the data inside the array needs to be changed dynamically. For e.g., when the instruction sequences are longer than the capacity of this memory, which can only hold 256 instructions. Writing into this memory dynamically is made possible by sharing address and data bus of the array. 
\end{itemize}

To identify the size of this memory for our architecture, we wrote the sequences for common operations like fixed-point addition, multiplication and MAC, and floating-point addition, multiplication and MAC. We found that the none of the operations was more than 200 instructions. So, we provide space for 256 instructions. With each instruction being 16-bit wide, the size of this memory is 4 Kilobits.



\subsubsection{Controller}
The controller in the Compute RAM block is required to fetch, decode, and execute the instructions in the instruction memory. It is implemented as a simple pipelined processor. The main array serves as the data memory for this processor. The number of registers in the register file is 8. In the sequences of common operations we wrote, we never used more than 5 registers at a time. The register file is implemented using flip-flops instead of a RAM to save area and allow multiple registers to be accessed at the same time. 

From the viewpoint of this controller, the instructions are of two types:
\begin{itemize}
    \item Instructions executed by the controller's execution unit. For example, branch or add a value to a register.
    \item Instructions sent to the main array. For example, performing bitwise add operation on bits stored in two rows in the main array.
\end{itemize}

The controller has a very simple execution unit - it only has 1 adder, 1 comparator and 1 logical unit. It does not have complex blocks like multipliers. We note that common DL operations involve repetitive instructions requiring loops. To reduce the number of instructions for an operation, the controller employs zero-overhead branch processing using dedicated hardware loop control, like in conventional DSP processors \cite{dsp_book}.

\subsubsection{Logic Peripherals}
Logic peripherals are present for each bitline (each column) 
as in \cite{neural_cache}. The sense amplifiers sense the result from two cells A and B, in the same bitline. BL gives $A.B$, while BLB gives $\bar{A}.\bar{B}$. 
To support floating point operations, we need instruction execution predicated on multiple conditions like the sign of a previous result. We add a 4-to-1 mux that selects the predication condition from among \texttt{Carry}, \texttt{NotCarry} and \texttt{Tag}. 


\subsection{Interface and Operation}

Table \ref{tab:interface} presents the name and description of the ports on a Compute RAM block. Most of the ports are the same as a BRAM (\texttt{address}, \texttt{data\_in}, \texttt{write\_en}, \texttt{data\_out}). We add some additional ports to the block to enable Compute RAM functionality (\texttt{start}, \texttt{mode}, \texttt{done}). Only 3 additional ports are added, minimizing the area, delay and routing overhead of adding new ports to an FPGA block.


\begin{table}[t]
  \renewcommand{\arraystretch}{1.1}
  \centering
  \caption{I/O interface of a Compute RAM block}
  \label{tab:interface}
  \footnotesize
  \begin{threeparttable}
  \begin{tabular}{
  | >{\centering\arraybackslash}m{0.15\columnwidth} 
  | >{\centering\arraybackslash}m{0.15\columnwidth}
  | >{\centering\arraybackslash}m{0.45\columnwidth}
  |}
    \hline
    \rowcolor{LightBlue} \centering
    \textbf{Signal} & \textbf{Direction} & \textbf{Function} \\
    \hline
    \texttt{mode}  & Input & Compute mode or storage mode \\
    \hline
    \texttt{start} & Input & Start executing instructions \\
    \hline
    \texttt{address} & Input & Read/write address \\
    \hline
    \texttt{data\_in} & Input & Write data \\
    \hline
    \texttt{write\_en} & Input & Read or write \\
    \hline
    \texttt{data\_out} & Output & Read data \\
    \hline
    \texttt{done}  & Output & Instruction execution finished \\
    \hline
  \end{tabular}
  \end{threeparttable}
\end{table}

The \texttt{mode} input specifies whether a user wants to use the Compute RAM block in compute mode or storage mode. In storage mode (\texttt{mode=0}), the block works exactly like a BRAM on an FPGA. The controller and logic peripherals as well as both \texttt{start} and \texttt{done} signals are not used in this mode. 
The instruction memory can be used as a regular BRAM by the application. The non-utilized structures has area overhead which is insignificant ($\sim$12\%).
    

In a typical use case, logic external to the Compute RAM (eg. a state machine implemented in LBs) will configure the Compute RAM in storage mode first. The input data will then be loaded into the array (e.g. from external DRAM). Then, the mode will be changed to compute mode and the \texttt{start} signal will be asserted. Instructions in the instruction memory will execute in order. When the last instruction (signalled by the presence of \texttt{end} instruction) is executed, the \texttt{done} signal is asserted. The external logic will wait for assertion of \texttt{done} before reading the results.

\subsection{Advantages and Limitations}
Computations are performed using the following resources on a baseline FPGA:
\begin{itemize}
    \item BRAM(s) to store the input operands and results
    \item LBs implementing control logic to orchestrate the data transfer and computation
    \item DSP slices or LBs to perform the actual computation
\end{itemize}


Compute RAM, on the other hand, provides the storage, the computation capability and the control logic integrated into one block. There are many advantages of using Compute RAMs:

\begin{enumerate}
    \item Because the computation happens inside the memory block, no wire and switching energy is spent in sending data to/from the compute units. Data movement between various blocks on the FPGA is significantly reduced. This leads to reduction in power consumption and an increase in energy efficiency. Another impact of the reduced dependence on the FPGA interconnect is that designs can now operate at higher frequencies of operation, thereby resulting in speeding up applications.
    \item Any custom operation with any custom precision can be supported by a Compute RAM block. No hardware with hardcoded support for specific operations and fixed number of precisions is involved in a Compute RAM. 
    For performing a different operation or for using a different precision, the instruction sequence needs to be modified. This can be done either at FPGA configuration time or at execution time. Changing the instruction sequence at execution time makes Compute RAMs programmable in a software-like manner. 
    \item Using a Compute RAM reduces the dependence on routing/interconnect resources because the computation can be done in the RAM itself. This means that the frequency of operation of designs using Compute RAMs can be higher than those using LBs, DSPs and RAMs. Hence, DL applications can be sped up by using Compute RAMs.
    \item Using Compute RAMs leads to reduced area to implement a given circuit. In comparison to a BRAM, a Compute RAM has an area overhead of the instruction memory, controller and peripheral logic. However, this area overhead is smaller than using a BRAM, a DSP slice and several LBs for realizing a computation on a baseline FPGA. This also means that larger circuits can now fit on the same FPGA chip. Adding Compute RAMs to FPGAs leads to an increase in the compute density of the FPGA $(GOPS/mm^2)$.  
\end{enumerate}

There are some limitations of adding Compute RAMs to FPGAs. Adding a new block on an FPGA means more heterogeneity make mapping/synthesizing harder. But all BRAMs can be replaced with Compute RAMs, thereby preserving the heterogeneity that exists today. Also, for some operations like floating point operations, Compute RAMs utilize some rows to store temporary results, reducing the overall capacity of the array. But these rows can be reused across all computations in a column and can be repurposed dynamically. Adding Compute RAMs to FPGAs means that users have to adopt a different programming model (writing instruction sequences), but this can be made easy by designing compilers and/or creating libraries of common operation sequences.

%
\section{Experimental Methodology} \label{meth}
\subsection{Tools} \label{tools}
We used the following tools to perform the experiments described in this paper:
\begin{itemize}
    \item VTR 8.0 for FPGA architecture exploration \cite{vtr8}
    \item Synopsys VCS 2018 for Verilog simulations \cite{synopsys_vcs}
    \item Synopsys Design Compiler 2018 for ASIC synthesis \cite{synopsys_dc}
    \item OpenRAM for estimating area and delays of SRAMs \cite{open_ram}
    \item COFFE 2.0 for estimating area and delays of FPGA components \cite{coffe2}
\end{itemize}

VTR is an academic tool for exploration of FPGA architectures. It takes two inputs - an FPGA architecture description file and a Verilog design file. In the FPGA architecture description file, the information of FPGA's building blocks and interconnect resources is provided. The Verilog design file contains the circuit we intend to map onto the FPGA. VTR synthesizes and implements the benchmark Verilog design for a hypothetical FPGA with the given architecture, and generates area and timing reports. 

\subsection{FPGA Architecture} \label{fpga_arch_eval}
For the experiments in this paper, we use an architecture similar to Intel Agilex \cite{intel_agilex} used by Arora et al. in \cite{tensor_slice_paper} as the baseline FPGA. Some of the properties of this FPGA architecture are as follows:
\begin{itemize}
    \item \textbf{Logic Block:} The logic block contains 10 basic logic elements. Each logic element consists of fracturable 6-input LUT, a flip-flop, and 2 bits of arithmetic. There are 60 inputs and 40 outputs on a logic block. 
    \item \textbf{DSP Slice:} The DSP slice supports addition (floating point only), multiplication, and MAC operations, along with some complex modes like $a*b+c$ or $(a+b)*c$. The precisions supported are 9-bit, 18-bit and 27-bit fixed point, and 16-bit (IEEE half precision and bfloat) and 32-bit (IEEE full precision) floating point.
    \item \textbf{BRAM:} The BRAMs are 20 Kilobits in size and can be configured as 512x40, 1024x20 and 2048x10. Both single port and dual port modes are supported.
    \item \textbf{Interconnect:} The routing channel width is 320. Wire segments of length 4 and 16 are used. The switch block is a Wilton Switch box with a flexibility of 3.
\end{itemize}

To this baseline FPGA architecture, we add Compute RAM blocks to create the proposed FPGA architecture. We evaluate the area and delay parameters of a Compute RAM block and plug in the description of a Compute RAM block in the FPGA architecture file. To find the area, we first obtain the BRAM area using \cite{coffe2}. We evaluate the area and delay of a 4 Kilobit RAM (instruction memory) using OpenRAM \cite{open_ram}. For the controller, we develop a simple pipelined processor in Verilog. We also design a logic peripheral block in Verilog. We then use Synopsys DC to synthesize these units and add a 15\% overhead of placement and routing \cite{domain_specific_fpgas}. Then, we add the area of a BRAM, instruction memory, controller and logic peripherals. The area overhead of adding 3 additional ports is negligible, so we ignore it. 
All the areas and delays in our results are based on the 22nm technology node. In some cases, because of unavailability of 22nm standard cell libraries, we used the 45nm GPDK library from Cadence, and scale the delays and areas based on equations present in \cite{Stillmaker201774}.

To evaluate the frequency of Compute RAM in compute mode, we identify the impact of modifications done to the BRAM on it's critical path delay. From \cite{supreet_logic_in_memory}, we see that the logic mode of the logic-in-memory RAM runs at $\sim$33\% reduced frequency compared to the storage mode, mainly due to the reduced voltage requirement and performing read \& write in the same cycle. We add to this the impact of adding logic peripherals ($\sim$3\%) to obtain the frequency of operation of Compute RAM. Note that the frequency of the Compute RAM in storage mode is not affected significantly and stays almost the same.

\subsection{Experimental Setup} \label{experiments}
The goal of our experiments is to evaluate the benefit of using Compute RAMs instead of baseline FPGA blocks (LBs, DSPs and BRAMs) for common operations like addition,  multiplication and dot product. We compare various metrics: area consumed, energy and total time taken. We use the most widely used precisions in FPGA DL accelerators: int4, int8 and bfloat16. However, it should be noted that Compute RAMs are  fully adaptable to any precision. 

The hardware designs used for the experiments include:
\begin{itemize}
    \item Memory to store the inputs and outputs.
    \item Compute units for performing the computation 
    \item Control logic to coordinate movement of operands and results between compute units and memory. 
\end{itemize}

In case of a baseline FPGA, we assume the design contains 1 BRAM (20 Kbits in 512x40 geometry) and that the data is laid out in the BRAM in the most optimal way to ensure maximum bandwidth usage. Compute units are LBs in case of fixed-point addition, and DSPs in other cases. We instantiate enough compute units to saturate the bandwidth from 1 BRAM. For example, for int4 addition operation, one row contains 3 input-output tuples (operand1, operand2, result), one row is read out in 1 cycle and the data is fed to 3 adders. For bfloat16 multiplication operation, three rows contain the operands and the results of 2 operations (row1 $\rightarrow$ \{operand1, operand2\}, row2 $\rightarrow$ \{operand3, operand4\}, row3 $\rightarrow$ \{result1, result2\}). Only 1 bfloat16 adder is enough to saturate the bandwidth provided by the BRAM. This is the most optimal configuration and ensures a fair comparison.

In case of the FPGA with Compute RAMs, most of the design is absorbed in a Compute RAM block. A Compute RAM block with 20 Kilobits capacity in the main array, with a geometry of 512x40 is used. We assume that the data is laid out in transpose format in the main array.

We run VTR with these designs and the baseline and proposed architectures to observe the area, delay/frequency and routing wirelength metrics.  We run VTR without a target frequency, which means it finds the fastest implementation possible. We disable any I/O to register and register to I/O paths from timing analysis.

On a baseline FPGA, the total cycles for an operation to complete include the time taken to read the inputs, perform the computation and the write the results. In the case of Compute RAM based FPGA, the total cycles for an operation to complete are the cycles to execute all the instructions in the instruction memory for a given operation. 

For energy, we add transistor energy and wire energy. For transistor energy, we use an activity factor of 0.1 and calculate the energy based on the number of transistors in each block (obtained from the area consumed by the block). For wire energy, we use wire energy numbers (fJ/mm/bit) from \cite{gpus_parallel_computing}, scale them to 22nm technology node and multiply them with the number of bits used for data transfer and the average net length obtained from VTR.

\section{Results} \label{results}

\subsection{Properties of Compute RAM}
Table \ref{tab:cram_vs_bram_vs_dsp} compares the various properties of a Compute RAM block with a DSP slice, a BRAM and a Logic block.
We observe that a Compute RAM has $\sim$33\% more area compared to a BRAM. The additional overhead comes from the existence of components like the instruction memory, controller and peripheral logic. A DSP Slice has $\sim$12\% more area than a Compute RAM block.

Compute RAMs are $\sim$37\% slower than BRAMs because of the lower voltage requirement for logic mode operation. But they are $\sim$43\% faster than DSPs in fixed-point mode and $\sim$67\% faster than DSPs in floating-point mode. DSP slice is slow even though it is pipelined because it is a large block with many I/O ports and has a large input crossbar in it. Compute RAMs are smaller than DSP slices and have a smaller number of inputs compared to a DSP slice leading to a smaller local input crossbar and hence shorter delay. The path delay through the main array of the Compute RAM is shorter than the combinatorial delay through a DSP as well. The frequency of operation of a Logic block varies with the size and nature of the computation or logic mapped to it.

The compute throughput per block (in giga operations per second (GOPS)) for different precisions can also be seen in the table. On a baseline FPGA, fixed-point additions are mapped to LBs, whereas other computations are mapped to DSPs. Mapping additions to DSPs is inefficient because of the lower frequency. Low-precision fixed-point multiplications (like int4) can be mapped to either LBs or DSPs efficiently. Floating-point operations are very inefficient when mapped to LBs. Compute RAMs, on the other hand, can be used for all computations, with higher throughput obtained for lower precisions. BRAMs are only used for storage, so their compute throughput is 0. 

\begin{table}[t]
  \renewcommand{\arraystretch}{1.1}
  \centering
  \caption{Comparison of Compute RAM, DSP, BRAM, and LB}
  \label{tab:cram_vs_bram_vs_dsp}
  \footnotesize
  \begin{threeparttable}
  \begin{tabular}{
  | >{\centering\arraybackslash}m{0.06\columnwidth}
  | >{\centering\arraybackslash}m{0.15\columnwidth} 
  | >{\centering\arraybackslash}m{0.13\columnwidth}
  | >{\centering\arraybackslash}m{0.17\columnwidth}
  | >{\centering\arraybackslash}m{0.08\columnwidth}
  | >{\centering\arraybackslash}m{0.11\columnwidth}
  |}
    \hline
    \rowcolor{LightBlue} 
    \multicolumn{2}{|c|}{\textbf{Metric}} & \textbf{Compute RAM} & \textbf{DSP Slice} & \textbf{BRAM} & \textbf{Logic Block} \\
    \hline
    \multicolumn{2}{|c|}{Area ($\mu m^2$)}  & 11072.5 & 12433 & 8311  & 1938 \\
    \hline
    \multicolumn{2}{|c|}{Frequency (MHz)} & 609.1 & 391.8 (fixed) 336.4 (float) & 922.9 & Varies \\
    \hline
    & Add (int4)  & 4.87    & -  & 0 & 2.2 \\
    \hhline{*{1}{|>{\arrayrulecolor{white}}-}>{\arrayrulecolor{black}}*{5}{-}}
    & Add (int8)  & 2.71    & -  & 0 & 0.61 \\
    \hhline{*{1}{|>{\arrayrulecolor{white}}-}>{\arrayrulecolor{black}}*{5}{-}}
    & Add (bf16)  & 0.31    & 0.22  & 0 & - \\
    \hhline{*{1}{|>{\arrayrulecolor{white}}-}>{\arrayrulecolor{black}}*{5}{-}}
    & Mul (int4)  & 1.21 & 0.52  & 0 & 0.21 \\
    \hhline{*{1}{|>{\arrayrulecolor{white}}-}>{\arrayrulecolor{black}}*{5}{-}}
    & Mul (int8)  & 0.34 & 0.52  & 0 & - \\
    \hhline{*{1}{|>{\arrayrulecolor{white}}-}>{\arrayrulecolor{black}}*{5}{-}}
    \multirow{-6}{*}{\rotatebox[origin=c]{90}{\makecell{Throughput \\ (GOPS)}}} & Mul (bf16)  & 0.27 & 0.22  & 0 & - \\
    \hline
  \end{tabular}
  \end{threeparttable}
\end{table}

\subsection{Addition}
\begin{figure}[hbt]
\centering
\includegraphics[width=0.8\linewidth]{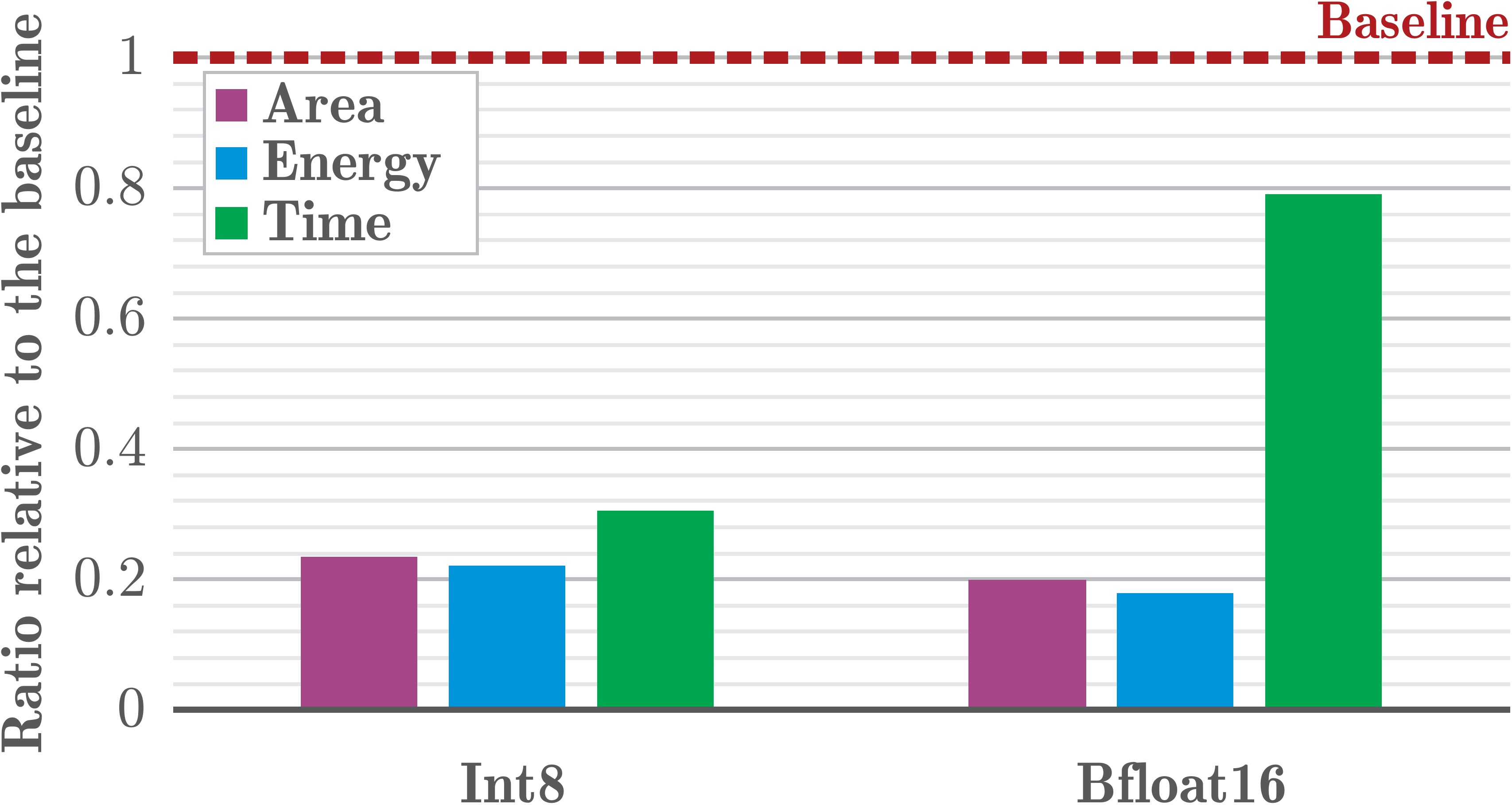}
\caption{Comparing a baseline FPGA with an FPGA with Compute RAMs for addition operation (RAM arrays are 512x40)}
\label{fig:addition_results}
\vspace{-2mm}
\end{figure}

Figure \ref{fig:addition_results} shows the results for addition operation. We compare various metrics for a baseline FPGA against an FPGA with Compute RAMs. The total number of addition operations in both cases are such that 20 Kilobits is required for storing all the operands and the results. 

Area consumed is the total areas of all the blocks (LBs, DSPs, BRAMs, Compute RAMs) used by the circuit on the FPGA. We observe significant reduction in area for both precisions. This is because in a baseline FPGA, soft logic (multiple LBs) is used for designing the control logic, but in case of Compute RAMs, the controller is hardened. 

The energy metric shows the dynamic energy consumed by the circuits mapped to the baseline FPGA and the FPGA with Compute RAMs.  We see that energy consumed when using Compute RAMs is $\sim$20\% of the energy consumed on baseline FPGA. This is because of the much lower dependence on FPGA interconnect fabric and also the reduced circuit area.

\vspace{-1mm}

The overall time taken shown in Figure \ref{fig:addition_results} is the product of cycles taken for the entire operation and the frequency of the circuit. 
The frequency of operation of the circuit is reported by VTR. 
The frequency of operation is 60-65\% higher when using Compute RAMs. This is because when using a baseline FPGA, there are paths between multiple DSPs, LBs and BRAMs through the interconnect fabric. These paths tend to be long and circuitous. When using Compute RAM, a very few short timing paths exist outside the Compute RAM.




For int8 precision, we see a significant reduction in time taken, because the number of cycles taken by Compute RAM is lower than the cycles taken on the baseline FPGA. However, for \texttt{bfloat16}, the time taken is only 20\% smaller. The number of cycles taken by Compute RAM is indeed larger in this case, because floating point addition requires a lot of steps. However, the overall time is still lower because the frequency of operation is much higher for Compute RAM.

\subsection{Multiplication}
\begin{figure}[t!]
\centering
\includegraphics[width=0.8\linewidth]{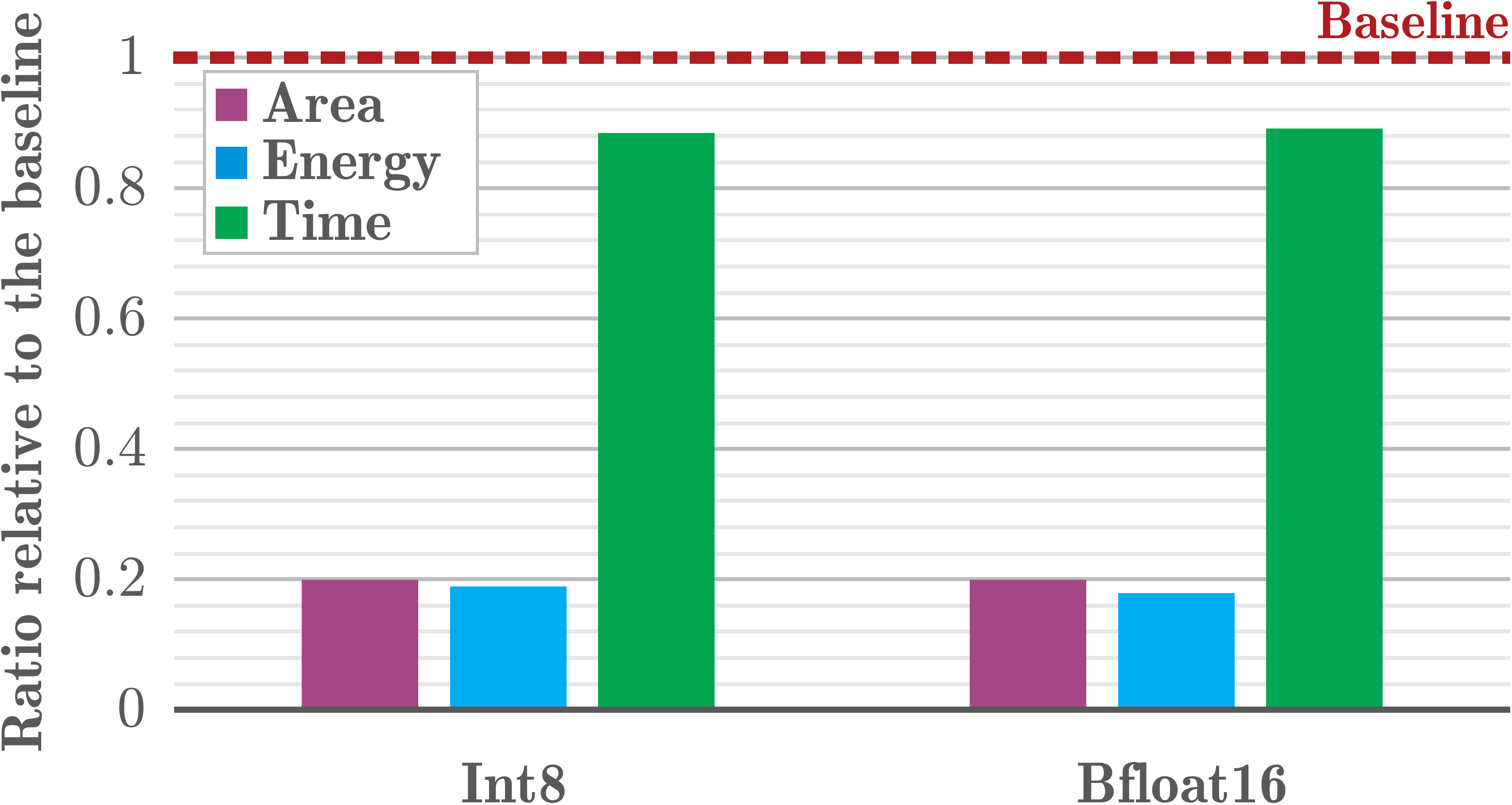}
\caption{Comparing a baseline FPGA with an FPGA with Compute RAMs for multiplication operation (RAM arrays are 512x40)}
\label{fig:multiplication_results}
\vspace{-2mm}
\end{figure}

Figure \ref{fig:multiplication_results} shows the results for multiplication operation. The total number of multiplication operations is such that 20 Kilobits is required for storing all the operands and the results. 

The area and energy results for multiplication are very similar to addition. 
The total time taken for multiplication operations is $\sim$12\% shorter for Compute RAMs than the baseline FPGA. Because of the bit-serial nature of the computation done by Compute RAMs, the number of cycles taken for multiplications is quite high. However, the overall time is still lower because the frequency of operation is much higher for Compute RAM.

\vspace{-2mm}

\subsection{Dot Product}
Dot product operation is the building block of neural networks. It is used in matrix-matrix multiplication and matrix-vector multiplication, which form 80-90\% of all computations in modern neural networks. Layers such as fully connected, convolution and LSTM are all based on these operations. Many FPGA-based hardware accelerators, ASIC chiplets and FPGAs have dot product engines in them \cite{brainwave} \cite{boutrosbeyond} \cite{TensorTile}. Dot product operation involves MAC and reduction operations. Two vectors are multiplied element-wise and the products are added to produce a scalar output.

In this section, we show the results of a dot product operation using int4 precision. The accumulation is performed using 32-bits (typical for DL). We consider vector sizes that ensure maximum utilization of the Compute RAM and the BRAM on a normal FPGA. On a baseline FPGA, there are 5 multipliers and 4 adders for accumulation, to ensure bandwidth provided by the BRAM is fully utilized.

\begin{figure}[t]
\centering
\includegraphics[width=0.8\linewidth]{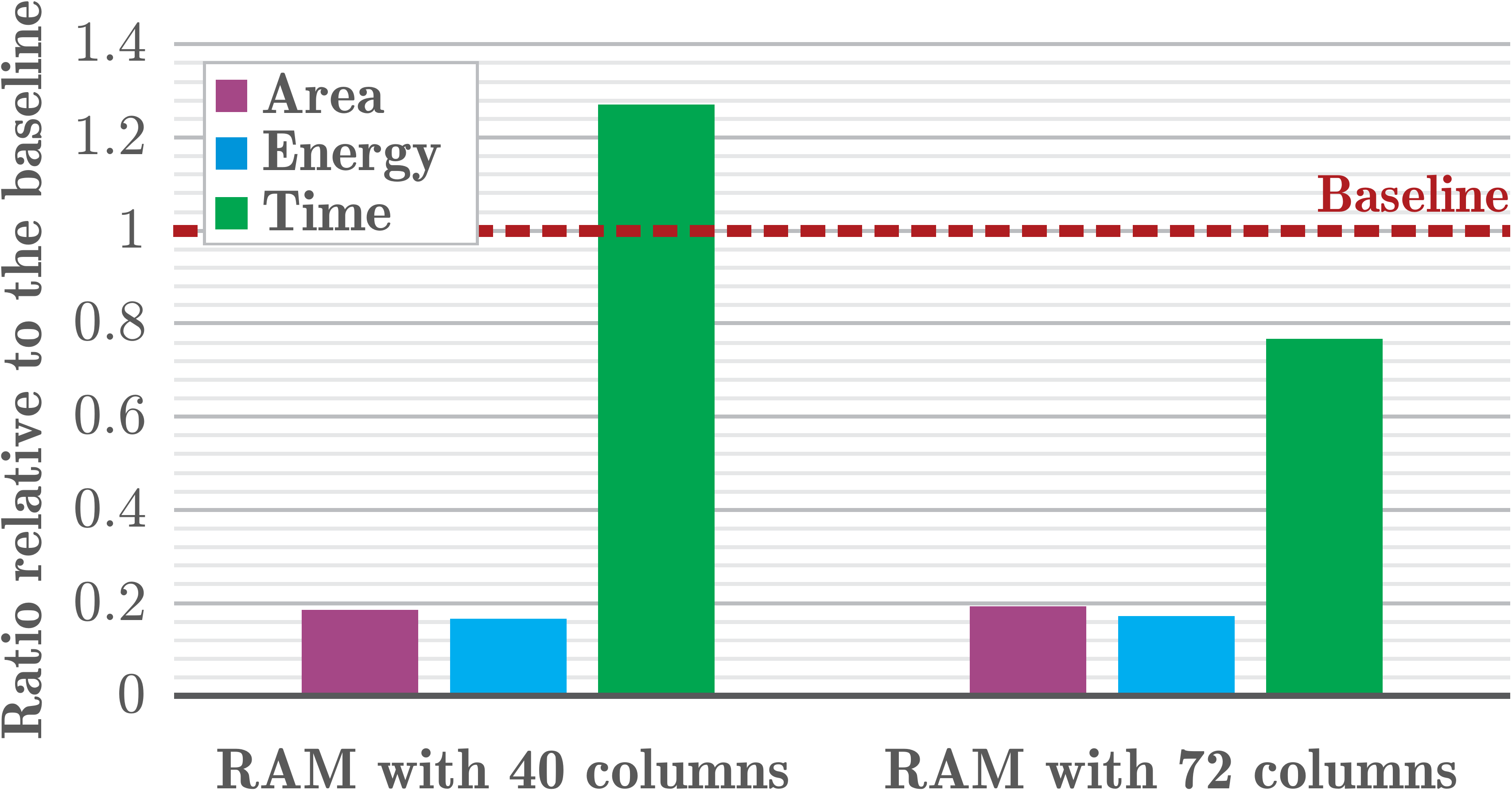}
\caption{Comparing a baseline FPGA with an FPGA with Compute RAMs for dot product operation in int4 precision}
\label{fig:dot_product_results}
\vspace{-2mm}
\end{figure}

The area and energy results are similar to the addition and multiplication results. However, there is an interesting observation pertaining to the overall time taken (referring to the left half of Figure \ref{fig:dot_product_results}). Compute RAM takes more time, even with the frequency of operation being higher. This is because Compute RAM takes much larger number of cycles compared to the baseline FPGA (1470 vs 480). In the design implemented on the baseline FPGA, there is enough parallelism. There are 5 4-bit multipliers running in parallel and 4 int32 adders perfoming accumulation in a tree structure. On top of that, all these compute units are pipelined. In Compute RAM, however, the parallelism is limited to the number of columns (bit lines), which is 40. To ensure maximal data packing and utilization of the Compute RAM, we store multiple input data items in one column. But within a column, the operations are performed serially. And the number of serial operations is relatively high in this case.

To reduce the time and improve the performance of Compute RAMs, more parallelism is required. This implies that a shallower and wider memory array will be required. BRAMs in Xilinx FPGAs have wider configurations up to 72 columns \cite{xilinx_ultrascale_bram}. We experiment with using 72 columns. We evaluate the impact of increasing the columns analytically and show it in the right half of Figure \ref{fig:dot_product_results}. We observe that there is minor impact on the area and energy metrics, but the total time is now $\sim$20\% better than the baseline (because of almost 2x the parallelism). Even more parallelism and speedup can be achieved if we increase the number of columns even further (say 40 rows x 512 columns, instead of 512 rows x 40 columns), but such a memory array will be expensive because large number of I/O ports, leading to significant changes in the interconnect architecture of the FPGA. We leave further detailed investigation of this topic as future work.

\section{Conclusion} \label{conclusion}
This paper proposes adding blocks called Compute RAMs to FPGAs to improve their performance for DL applications. A Compute RAM block enables processing-in-memory by utilizing an emerging bit-line SRAM circuit technology coupled with bit-serial arithmetic. Each individual operation is performed serially, but multiple operations are done in parallel in the same block. This unlocks performance benefits for parallel throughput-oriented compute-intensive operations which tend to involve a lot of on-chip data movement if implemented on current FPGAs using Logic Blocks, DSP slices and BRAMs. 

We present the architecture of Compute RAM blocks, propose adding them to FPGAs, and describe how computations can be orchestrated using them. We demonstrate the efficacy of these blocks for common DL math operations. We believe that Compute RAMs can replace BRAMs on existing FPGAs, and transform them into massively parallel computation units, while still performing the traditional role of acting as storage units. Compute RAMs can work in tandem with DSP slices and logic blocks, leading to a many fold enhancement in the throughput and compute density of FPGAs. In the future, we plan to evaluate the performance boost that can be obtained at the application level (neural networks) by using these blocks. We also plan to explore using shallower, wider RAMs to increase the amount of parallelism and speedup.



\section*{Acknowledgement}
\noindent
This research was supported in part by National Science Foundation (NSF) grant number 1763848. Any opinions, findings, conclusions or recommendations are those of the authors and not of the NSF.

\vspace{-0.2cm}
\bibliographystyle{IEEEtran}
\bibliography{bibliography}

\end{document}